\documentclass[12pt]{article}
\usepackage{graphicx}
\usepackage{cite}
\usepackage{url}
\usepackage[utf8]{inputenc}
\usepackage{hyperref}
\usepackage{fancyvrb}
\usepackage{etoolbox}

\newcommand\pubnumber{}
\newcommand\pubdate{March 18, 2016}

\def\kansas{Department of Physics and Astronomy\\
University of Kansas, Lawrence, KS 66045, U.S.A.}
\def\support{\footnote{Work supported by the National Science Foundation  
                       under award PHY-1306953.}}

\def\Title#1{\begin{center} {\Large #1 } \end{center}}
\def\Author#1{\begin{center}{ \sc #1} \end{center}}
\def\Address#1{\begin{center}{ \it #1} \end{center}}

\newcommand\pubblock{\rightline{\begin{tabular}{l} \pubnumber\\
         \pubdate  \end{tabular}}}
\newenvironment{Abstract}{\begin{quotation}  }{\end{quotation}}
\newenvironment{Presented}{\begin{quotation} \begin{center} 
             Talk presented at \end{center}\bigskip 
      \begin{center}\begin{large}}{\end{large}\end{center} \end{quotation}}
\def\Acknowledgments{\bigskip  \bigskip \begin{center} \begin{large}
             \bf Acknowledgments \end{large}\end{center}}

\def\beq{\begin{equation}}
\def\eeq#1{\label{#1}\end{equation}}
\def\eeqn{\end{equation}}

\def\beqa{\begin{eqnarray}}
\def\eeqa#1{\label{#1}\end{eqnarray}}
\def\eeqan{\end{eqnarray}}

\let\bar=\overbar

\def\Dslash{\not{\hbox{\kern-4pt $D$}}}
\def\dslash{\not{\hbox{\kern-2pt $\del$}}}

\def\msb{{\bar{\ssstyle M \kern -1pt S}}}

\RecustomVerbatimCommand{\VerbatimInput}{VerbatimInput}%
{fontsize=\footnotesize,
 frame=lines,  
 framesep=2em, 
 rulecolor=\color{gray},
 label=\fbox{\color{black}},
 labelposition=topline,
 commandchars=\|\(\) 
}

\begin{document}
\begin{titlepage}
\pubblock

\vfill
\Title{Updated Study of a Precision Measurement of the W Mass from 
a Threshold Scan Using Polarized $\rm{e}^-$ and $\rm{e}^+$ at ILC}
\vfill
\Author{GRAHAM W. WILSON\support}
\Address{\kansas}
\vfill
\begin{Abstract}
An updated study of measuring the W mass from a polarized threshold scan at ILC is presented with
an emphasis on evaluating scan strategies that control experimental systematics.
Highly longitudinally polarized beams of electrons and positrons 
such as are feasible at ILC offer significant advantages in terms of statistical power 
and in-situ control of background.
Eventual experimental precision of around 2 MeV can be envisaged from this technique.
Further work on both the accelerator design and theoretical uncertainties
will likely be needed to take full advantage of this opportunity.
\end{Abstract}
\vfill
\begin{Presented}
The International Workshop on Future Linear Colliders (LCWS15),\\
Whistler, Canada, 2-6 November 2015\\
\end{Presented}
\vfill
\end{titlepage}
\def\thefootnote{\fnsymbol{footnote}}
\setcounter{footnote}{0}

\AtEndEnvironment{thebibliography}{
}

\section*{Introduction}

A future high energy $\rm{e}^+\rm{e}^-$ collider is recognized as 
essential for a precision study of the Higgs and the top quark~\cite{Baer:2013cma}. 
It can also be a very powerful tool for advancing 
measurements of precision electroweak 
observables~\cite{Baer:2013cma,Moortgat-Picka:2015yla,Freitas:2013xga,Baak:2013fwa}. 
One of those observables of considerable importance is the W mass. Measurements 
from LEP2 and the Tevatron have led to a current precision of 15~MeV~\cite{PDG-2014}.
Further improvements from long existing hadron collider data-sets at the Tevatron and LHC are possible, 
but given the predominant systematic uncertainties will constitute major experimental 
and phenomenological tours de force if and when they are realized.

The three most promising approaches to measuring 
the W mass at an $e^{+} e^{-}$ collider are: 
\begin{description}
\item[Polarized Threshold Scan] Measurement of  
the $\mathrm{W}^+\mathrm{W}^-$ cross-section near threshold with longitudinally 
polarized beams.
\item [Constrained Reconstruction] Kinematically-constrained reconstruction of $\mathrm{W}^+\mathrm{W}^-$ using 
constraints from four-momentum conservation and optionally mass-equality 
as was done at LEP2.
\item [Hadronic Mass] Direct measurement of the hadronic mass. This can be applied 
particularly to single-W events decaying hadronically 
or to the hadronic system in semi-leptonic $\mathrm{W}^+\mathrm{W}^-$ events.
\end{description}
Methods for measuring the W mass in $\rm{e}^+\rm{e}^-$ colliders were explored extensively in the LEP era, 
see~\cite{Kunszt,Stirling} and references therein.

The International Linear Collider (ILC) is designed to 
reach $\sqrt{s} \approx 500$~GeV with polarized beams and high luminosity and can 
be upgraded to $\sqrt{s} \approx 1$~TeV. 
With the envisaged accelerator parameters and ILC operating scenarios~\cite{Barklow:2015tja}, 
there is a great potential for much improved measurements of the W mass. 
With the example operating scenario H-20 from~\cite{Barklow:2015tja}, 
one can envisage data-sets totaling up to 6200 fb$^{-1}$ at center-of-mass 
energies of between 250 and 500 GeV. 
This is an energy regime where the W mass can be measured using the 
constrained reconstruction and hadronic mass techniques. The 
eventual uncertainty would almost certainly be limited by experimental systematics. 
Previous rough estimates by the author in ~\cite{Baak:2013fwa} suggest experimental systematics 
on the 3-4 MeV scale which would dominate the statistical uncertainties.  
Nevertheless given the opportunity to improve on the measurement of the W mass using data collected 
synergistically with the main ILC physics program this is an area where further 
detailed study would be very welcome.

The main subject of this contribution is an update of a previous study on the measurement of the W mass 
using a polarized threshold scan~\cite{Wilson_Sitges, Wilson_TDR2}. 
The study has evolved taking into account additional systematic effects.
The updates include the use of beam parameters consistent with the ILC TDR design and 
experimental performance appropriate to the envisaged ILC detectors. The previous study was 
started in 1999 and had very conservatively assumed experimental characteristics similar to 
the LEP detectors. 

Such a measurement would necessarily entail significant allocation of running time to data-taking near $\sqrt{s}=161$~GeV where 
the cross-section is most sensitive to $M_{\mathrm{W}}$. 
Low statistics measurements at a single center-of-mass energy with unpolarized beams were done in 1996 
by the LEP experiments~\cite{ALEPH, DELPHI, L3, OPAL} and were reviewed in~\cite{Oxford-Primary,Oxford-Secondary}.
The dependence of the cross-section on center-of-mass energy is illustrated in Figure 1. 
There are two primary experimental issues at the heart of 
interpreting a high statistics threshold scan as a measurement of the W mass. Firstly, one needs excellent 
control of the {\bf absolute center-of-mass energy}, and secondly one needs to be able to control the {\bf background}.
These are discussed further in the next section.

\begin{figure}[!htb]
  \begin{center}
    \includegraphics[width=0.95\textwidth]{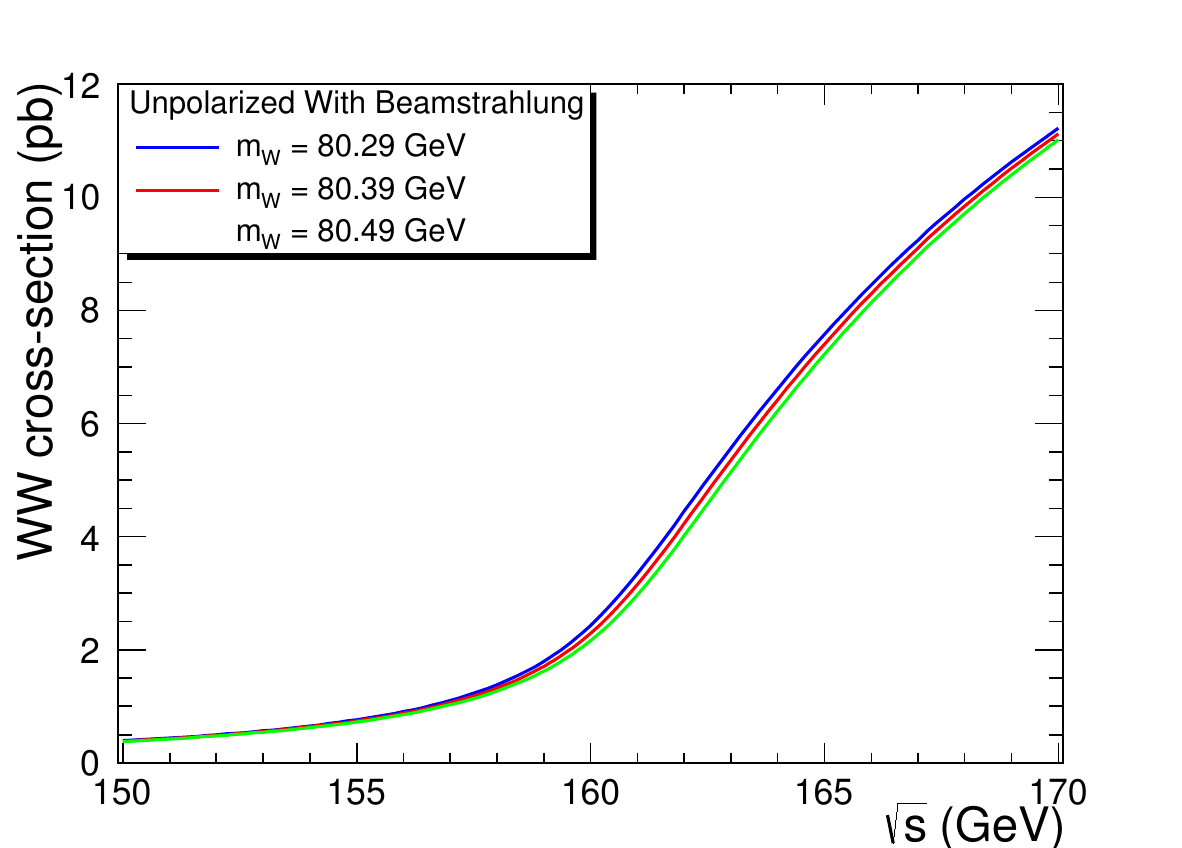} 
    \caption{The unpolarized CC03 cross-section for WW production vs center-of-mass energy. 
The cross-section is evaluated with GENTLE2.0 including ILC beamstrahlung.} 
    \label{fig:WWcrosssection}
  \end{center}
\end{figure}

\section*{Strategy for Primary Experimental Systematics}

It has been shown recently~\cite{Wilson-ECMP} that $\rm{e}^+\rm{e}^- \rightarrow \mu^+ \mu^- (\gamma)$ events can be used 
to make an {\it in situ} measurement of the average center-of-mass energy 
with high statistical precision\footnote{Previous studies 
had assumed that the absolute center-of-mass energy would be determined 
using the angle technique in Z$\gamma$ events (with $\rm{Z} \rightarrow \mu^+ \mu^-$) - a technique that 
suffers from relatively poor event-by-event statistical precision given the Z width} 
based simply on momentum measurements of the muons as first discussed in~\cite{Barklow-ECMP}.
Under the assumption that the recoil mass to the measured muons is zero, 
an estimate of the center-of-mass energy of such events can be formed simply from the measured momenta of the two muons:
\[ \sqrt{s}_P = E_1 + E_2 + |\vec{p_1} + \vec{p_2}| = \sqrt{p_1^2 + m_{\mu}^2} + \sqrt{p_2^2 + m_{\mu}^2} + |\vec{p_1} + \vec{p_2}|    \]
The distribution of this variable can then be used to deduce relevant parameters including those related to the 
average absolute center-of-mass energy.
Given very good control of the tracking detector absolute 
momentum scale\footnote{Assumed known to 10 ppm through momentum-scale calibrations using 
at least 12,000 energetic J/$\psi \rightarrow \mu^+ \mu^-$ events as outlined in~\cite{Wilson-JPsi}}, 
it is envisaged that knowledge of the absolute center-of-mass energy at the 10 ppm level can be targeted, corresponding to a 
contribution to the W mass uncertainty at threshold of 0.8 MeV.

Our strategy for controlling the background is to measure the background {\it in situ} using polarized beams. 
Polarized electrons and {\bf polarized positrons} make it feasible to measure the background 
and simultaneously measure the polarization (also {\it in situ}).
Having good control of the background is critical. Assuming a background level of 250~fb, a 10\% uncertainty on the 
background amounts to an uncertainty of 12~MeV on $M_{\mathrm{W}}$ for 
an unpolarized scan. With highly polarized beams, and a 100~fb$^{-1}$ polarized scan, 
the background level can be measured to about 4~fb. 

In addition to the need for exclusive running near WW threshold, the polarized threshold scan 
is most attractive if the beams are highly polarized, and if sufficient data-sets are collected to pin-down and 
to monitor the tracker momentum scale which affects the determination of the absolute center-of-mass energy. 
At present the most obvious way to guarantee the latter, is to make sure that the accelerator can also run 
effectively at the Z pole yielding high statistics calibration data.
In these three aspects there is a need for the ILC accelerator design to retain compatibility with 
good performance at relatively low energy.

\section*{Polarized Beams}
The cross-section dependence on the longitudinal polarization of the 
electron and positron beams is given by~\cite{MoortgatPick:2005cw} 
\begin{eqnarray*}
\sigma(P_{\mathrm{e^-}}, P_{\mathrm{e^+}}) & = &  \frac{1}{4} \{ 
(1-P_{\mathrm{e^-}}) (1+P_{\mathrm{e^+}}) \sigma_{LR} +
(1+P_{\mathrm{e^-}}) (1-P_{\mathrm{e^+}}) \sigma_{RL} + \\  
 & & \quad \; (1-P_{\mathrm{e^-}}) (1-P_{\mathrm{e^+}}) \sigma_{LL} +
(1+P_{\mathrm{e^-}}) (1+P_{\mathrm{e^+}}) \sigma_{RR}  
\}
\end{eqnarray*}
where $\sigma_{k}$ ($k=$ LR, RL, LL and RR)  are the fully polarized cross-sections.
In cases where the LL and RR cross-sections are zero, the resulting cross-section simplifies to
\begin{eqnarray*}
\sigma(P_{\mathrm{e^-}}, P_{\mathrm{e^+}}) & = &  \frac{1}{4} \{ 
(1-P_{\mathrm{e^-}}) (1+P_{\mathrm{e^+}}) \sigma_{LR} +
(1+P_{\mathrm{e^-}}) (1-P_{\mathrm{e^+}}) \sigma_{RL}
\}
\end{eqnarray*}
which can be rewritten as 
\begin{equation}
\sigma(P_{\mathrm{e^-}}, P_{\mathrm{e^+}}) = \sigma_u \{ (1 - P_{\mathrm{e^-}} \: P_{\mathrm{e^+}} ) - 
(P_{\mathrm{e^-}} - P_{\mathrm{e^+}}) A_{LR} \}
\end{equation}
where $\sigma_u$ is the unpolarized cross-section, ($\sigma_{LR}+\sigma_{RL}$)/4, and $A_{LR}$ is the left-right asymmetry defined as 
\[  A_{LR} = (\sigma_{LR} - \sigma_{RL}) / (\sigma_{LR} + \sigma_{RL})   \]
Equation 1 is appropriate for Z production. It is also appropriate for 
the doubly-resonant CC03 WW production diagrams, where especially close to threshold, $A_{LR}$, is close to maximal\footnote{The estimated 
value for WW is around 0.99. Checks with Wopper~\cite{Wopper} and RacoonWW~\cite{Racoon} gave values of 0.992 and 0.988 near threshold} 
given the dominance of the t-channel $\nu_{\mathrm{e}}$ exchange diagram.
For many of the processes, that play the role of a background to WW production, notably, $\rm{e}^+\rm{e}^- \rightarrow q \overline{q} g g$, it 
is also highly appropriate. Note that this is not appropriate for processes such 
as single W production which is expected to contribute 
to the background in the $q \overline{q} \mathrm{e} \nu$ channel (the LL and RR cross-sections are non-zero in this case).

With the WW asymmetry of around 0.99, and background asymmetries ranging from 0.15 to 0.48 (depending on channel), it is feasible 
to use the polarization of the beams to preferentially enhance the signal cross-section, and therefore the statistics for the measurement. 
Conversely, it is possible to use the polarization to essentially turn-off the signal process, and measure events that are 
much enriched in background.

At ILC, it is expected that the beams can be highly polarized, and the spin can 
be flipped with high frequency (certainly pulse-to-pulse for electrons).
Electron polarization is straightforward.
Electron polarizations of 80\% are in the baseline, and values as high as 90\% can be targeted. 
The baseline ILC design has polarized positron beams at beam energies exceeding 125 GeV with 
a polarization level of 30\%, and there are studies and prospects for positron polarization levels as high as 60\%. 
Fully worked out designs for high instantaneous luminosity and positron polarization at low energy are under study.
The relative polarization will be monitored with polarimeters.

Experimentally, we expect to have the freedom to choose appropriate fractions of the delivered luminosity in the various 
polarization configurations. We can imagine collisions with positive, negative and zero polarization for each beam, resulting 
in a total of 9 different ``helicity configurations'', namely ($-+$, $+-$, $--$, $++$, $00$, $0+$, $0-$, $+0$, $-0$).

\begin{table}[!htb]
\begin{center}
\begin{tabular}{l|r|c|c|r|r|r|r}
    $\sqrt{s}$ (GeV) &   L ($\rm{fb}^{-1}$)   &  $f$  & $\lambda_{\rm{e}^{-}} \lambda_{\rm{e}^{+}} $   &   $N_{ll}$ & $N_{lh}$   & $N_{hh}$ & $N_
{RR}$ \\ \hline 
     160.6 &      4.348 &     0.7789  & $-+$   &    2752  &   11279  &   12321  &  926968 \\ 
           &            &     0.1704  & $+-$   &      20  &      67  &     158  &  139932 \\ 
           &            &     0.0254  & $++$   &       2  &      19  &      27  &    6661 \\ 
           &            &     0.0254  & $--$   &      21  &     100  &     102  &    8455 \\ \hline

     161.2 &     21.739 &     0.7789  & $-+$   &   16096  &   67610  &   73538  & 4635245 \\ 
           &            &     0.1704  & $+-$   &      98  &     354  &     820  &  697141 \\ 
           &            &     0.0254  & $++$   &      37  &     134  &     130  &   33202 \\ 
           &            &     0.0254  & $--$   &     145  &     574  &     622  &   42832 \\ \hline

     161.4 &     21.739 &     0.7789  & $-+$   &   17334  &   72012  &   77991  & 4639495 \\ 
           &            &     0.1704  & $+-$   &     100  &     376  &     770  &  697459 \\ 
           &            &     0.0254  & $++$   &      28  &     104  &     133  &   33556 \\ 
           &            &     0.0254  & $--$   &     135  &     553  &     661  &   42979 \\ \hline

     161.6 &     21.739 &     0.7789  & $-+$   &   18364  &   76393  &   82169  & 4636591 \\ 
           &            &     0.1704  & $+-$   &      81  &     369  &     803  &  697851 \\ 
           &            &     0.0254  & $++$   &      43  &     135  &     174  &   33271 \\ 
           &            &     0.0254  & $--$   &     146  &     618  &     681  &   42689 \\ \hline

     162.2 &      4.348 &     0.7789  & $-+$   &    4159  &   17814  &   19145  &  927793 \\ 
           &            &     0.1704  & $+-$   &      16  &      62  &     173  &  138837 \\ 
           &            &     0.0254  & $++$   &      10  &      28  &      43  &    6633 \\ 
           &            &     0.0254  & $--$   &      46  &     135  &     141  &    8463 \\ \hline

     170.0 &     26.087 &     0.7789  & $-+$   &   63621  &  264869  &  270577  & 5560286 \\ 
           &            &     0.1704  & $+-$   &     244  &     957  &    1447  &  838233 \\ 
           &            &     0.0254  & $++$   &     106  &     451  &     466  &   40196 \\ 
           &            &     0.0254  & $--$   &     508  &    2215  &    2282  &   50979 \\ \hline
\end{tabular}
\caption{Illustrative example of the numbers of events in each channel for the 
standard 100~$\rm{fb}^{-1}$ 6-point ILC scan with 4 helicity configurations. Columns give the center-of-mass energy, $\sqrt{s}$, 
the apportioned integrated luminosity, the fraction for each helicity configuration, $\lambda_{\rm{e}^{-}} \lambda_{\rm{e}^{+}} $, and the numbers 
of events observed in each channel.}
\label{tab:illustrate}
\end{center}
\end{table} 

\section*{Example Scan}
The basic method is to do a counting experiment where the experimental observables are the number 
of selected candidate events, $N_{ijk}$,  
consistent with WW production in the different decay channels (index $i$), observed at each 
center-of-mass energy (index $j$) and for each helicity configuration of the polarized electron and positron beams (index $k$).
The data are sub-divided into the 3 major decay channels ($q\overline{q}q\overline{q}$, $q\overline{q} \ell \nu$, $\ell \nu \ell\nu$).
In addition to the selected candidate events, we also measure the 
number of Z-like $\rm{e}^+\rm{e}^- \rightarrow f \overline{f} (\gamma)$ events produced for each center-of-mass energy 
and helicity configuration, $N^Z_{jk}$, 
as a means of measuring the polarization {\it in situ}.
An example simulated data-set corresponding to a 6-point ILC scan with 4 helicity configurations, and correspondingly 96 event counts, 
is displayed in Table~\ref{tab:illustrate}.

The observed event counts in the various channels are fitted 
to the expected event counts from signal plus background events in a model with fit parameters that 
account for the theoretical model parameters and relevant systematic effects.
The fits are done using a Poisson likelihood.
The fit parameters are given in Table~\ref{tab:fitparameters}.

\begin{table}[!htb]
\begin{center}
\begin{tabular}{r|c|c}
No. & Fit Parameter & Comment \\ \hline
1 &  $m_{W}$  &  \\
2 &  $\alpha_S$ &  Fixed currently to 0.12 \\ \hline
3 &  $f_l$    &  0.1\% constrained \\   \hline
4 &  $\varepsilon$ (lvlv) & \\ 
5 &  $\varepsilon$ (qqlv) & Signal efficiency \\ 
6 &  $\varepsilon$ (qqqq) &  (constrained) \\  \hline
7 &  $\sigma_B$ (lvlv) &  \\ 
8 &  $\sigma_B$ (qqlv) & Background cross-section \\ 
9 &  $\sigma_B$ (qqqq) &  \\  \hline  
10 & $A_{LR}^B $ (lvlv) & \\
11 & $A_{LR}^B $ (qqlv) & Background asymmetry \\
12 & $A_{LR}^B $ (qqqq) &  (constrained) \\    \hline
13 & $\beta_B$ (lvlv) &  \\
14 & $\beta_B$ (qqlv) &  Background shape \\ 
15 & $\beta_B$ (qqqq) &  \\  \hline
16 & $|P(e^{-})|$  &  Assume same for each helicity  \\
17 & $|P(e^{+})|$  &  Assume same for each helicity  \\
18 & $\sigma_{\mathrm{Z}}$  &  Z-like 2-fermion ( $f \overline{f} (\gamma)$ ) \\
19 & $A_{LR}^{\mathrm{Z}} $ &  \\           \hline
\end{tabular}
\caption{19-parameter fit for $\ge$~4 helicity configuration scans ($-+$, $+-$, $++$, $--$)}
\label{tab:fitparameters}
\end{center}
\end{table} 

\section*{Scan Details}

The shape of the cross-section depends on $m_{W}$ and $\Gamma_{W}$. 
Within the Standard Model, $\Gamma_{W}$, is essentially a function 
of $m_{W}$ and $\alpha_S$.
\begin{equation}
\Gamma_W \sim {M_W^3} \left( 1 + \frac{2 \alpha_S( M_W^2) } { 3 \pi} \right) 
\end{equation}

The theoretical form of the WW CC03 cross-section is evaluated 
using GENTLE2.0/4fan~\cite{Bardin:1996zz}. 
The GENTLE predictions are convolved with the expected ILC beamstrahlung 
using CIRCE1~\cite{Ohl:1996fi}. The beamstrahlung spectrum was evaluated using Guinea-Pig~\cite{Schulte:1998au} 
by $\gamma$ scaling of the ILC TDR accelerator parameters~\cite{ILC_TDR_BASELINE} at $\sqrt{s}=200$~GeV to 161 GeV. 
The resulting energy loss function per beam is then fitted using the methods described in~\cite{Sailer:2009zz}, 
resulting in the four standard parameters of the Circe parameterization with 
values found of 0.70648, 0.25305, 50.507 and -0.7305.

The chosen theoretical model fit parameters are $m_{W}$ and $\alpha_S$. 
For the present studies, $\alpha_S$, was fixed to 0.12\footnote{Some details on studies related to $\Gamma_W$ sensitivity 
were discussed in~\cite{Wilson_TDR2}}.

\begin{table}[!htb]
\begin{center}
\begin{tabular}{l|c|c|c|c|c|c}
Channel   & Efficiency (\%) & $\sigma^U_{\mathrm{bkgd}}$ (fb) & $A_{\mathrm{LR}}^B$ & Eff. syst. (\%) 
& Bkgd syst.  & $A_{\mathrm{LR}}^B$ syst.\\ \hline
lvlv        &       87.5   &   10  & 0.15 &      0.1        &   free      & 0.025    \\
qqlv        &       87.5   &   40  & 0.30 &      0.1        &   free      & 0.012    \\
qqqq        &       83.5   &  200  & 0.48 &      0.1        &   free      & 0.005    \\ \hline
\end{tabular}
\caption{Experimental assumptions for the WW event selection near threshold using a polarized scan}
\label{tab:Assumptions}
\end{center}
\end{table} 

Other fit parameters related to the normalization and especially the normalization of the signal, include a scale factor for systematic 
uncertainty on the absolute integrated luminosity, $f_l$, and scale factors for 
corrections to the estimated efficiency in each channel, $\varepsilon$ (lvlv), $\varepsilon$ (qqlv), $\varepsilon$ (qqqq).
All four of these uncertainties are constrained within specified uncertainties. The constraint is accomplished by 
adding a $\chi^2$ penalty contribution.

Parameters 16-19 are used to measure the beam polarization using the Z-like events in a similar manner to~\cite{Blondel}. 
Using equation 1, we see that with 4 cross-section measurements, corresponding to, 
for example, the $-+$, $+-$, $--$, $++$ helicity configurations, it is possible to 
measure these four parameters, namely, $\sigma_u$, $A_{LR}$, $|P(e^{-})|$ and $|P(e^{+})|$. 
This assumes that the polarization magnitudes are identical for positive and negative polarization of the same 
beam (perfect spin-flip approximation). If this is not the case, or if 
this assumption needs to be checked it would be possible to take additional data with unpolarized beam or beams to further 
constrain the polarization model.

The other 9 parameters pertain to the background modeling in each of the 3 channels. 
The extraction of $M_W$ from the cross-section and its dependence on $\sqrt{s}$ near threshold 
is sensitive to the understanding of the background. 
It is expected that the background contribution needs to be  
determined from data in a robust way.
For each channel, the parameters, are the background cross-section, $\sigma_B$, 
the background left-right asymmetry, $A_{LR}^B $, and a background shape 
parameter, $\beta_B$, allowing for a power-law center-of-mass energy dependence of 
the background cross-section according to 
\[   \sigma(\sqrt{s}) = \sigma_B \left( \frac{161}{\sqrt{s}} \right)^{\beta_B} \]
Reasonable guesses for the shape parameter $\beta_B$ are in the range [-2,2]. 
This is expected to be a reasonable model for the qqqq channel where background issues 
are a major concern. Input data-sets use $\beta_B=0$. 

With polarized beams, there is little need to take data at $\sqrt{s}$ values far 
from threshold to measure the background, but there is a need to measure the 
polarization and to control the polarization systematics of the background. 
It is assumed that the background asymmetries can be constrained with background ``side-bands'', and these parameters 
are also implemented with a $\chi^2$ penalty function.
The standard fits 
discussed here with both beams polarized only use the first six background parameters.
Other cases of interest are with only electron beam polarization, where one may need to rely more 
on external measurements of the beam polarization, and some data-taking below threshold.
Especially with no polarization, data-taking below threshold would appear 
mandatory for background control, but it would also be important to understand the $\sqrt{s}$ 
shape dependence of the background, and the $\beta_B$ parameters have been introduced to start to address this issue.

\section*{Polarized Threshold Scan Study}

The improvements include a re-optimization of the fraction of the luminosity 
associated with each beam helicity configuration which results from 
the assumed better detector performance.
The updated assumptions on the experimental event selection and the associated systematics are 
given in Table~\ref{tab:Assumptions}. These correspond to a factor of two reduction in the 
event selection inefficiency and a factor of two reduction in the non-WW backgrounds compared to 
that essentially achieved with the LEP detectors. Further improvement beyond these expected 
performance numbers is not out of the question.

For the current studies $A_{LR}^{WW}$ was set at 1.0. The cross-section 
for Z-like events was taken to be 150~pb with 
a value of $A_{LR}^Z$ of 0.19 reflecting a mix of full energy 
and radiative-return contributions.

\begin{table}[!htb]
\begin{center}
\begin{tabular}{c|c|c}
Fit parameter & Value & Error \\ \hline
   $m_{W}$ (GeV) &      80.388     &  3.77 $\times 10^{-3}$  \\ \hline
   $f_l$ &      1.0002     &  0.924 $\times 10^{-3}$  \\ \hline
   $\varepsilon$ (lvlv) &     1.0004     &  0.969 $\times 10^{-3}$  \\
   $\varepsilon$ (qqlv) &    0.99980     &  0.929 $\times 10^{-3}$  \\
   $\varepsilon$ (qqqq) &     1.0000     &  0.942 $\times 10^{-3}$  \\ \hline
   $\sigma_B$ (lvlv) (fb) &     10.28  &  0.92   \\
   $\sigma_B$ (qqlv) (fb) &     40.48  &  2.26  \\
   $\sigma_B$ (qqqq) (fb) &     196.37     &  3.62 \\
   $A_{LR}^B $ (lvlv) &     0.15637     &  0.0247  \\
   $A_{LR}^B $ (qqlv) &     0.29841     &  0.0119  \\
   $A_{LR}^B $ (qqqq) &     0.48012     &  4.72 $\times 10^{-3}$  \\ \hline
   $|P(e^{-})|$   &     0.89925     &  1.27 $\times 10^{-3}$  \\
   $|P(e^{+})|$  &     0.60077     &  9.41 $\times 10^{-4}$  \\
   $\sigma_{\mathrm{Z}}$ (pb) &      149.93     &  0.052  \\ 
   $A_{LR}^{\mathrm{Z}}$ &     0.19062     &  2.89 $\times 10^{-4}$  \\
\hline
\end{tabular}
\caption{Example fit of the 6-point ILC scan with 100 fb$^{-1}$ illustrated in Table~\ref{tab:illustrate}. In this example, 
the background $\beta_B$ shape parameters are fixed to zero, and $\alpha_S$ is fixed at 0.12.}
\label{tab:Minuit}
\end{center}
\end{table}

The re-optimized running strategy for 100 fb$^{-1}$ with 90\% $\rm{e}^-$ polarization and 60\% $\rm{e}^+$ polarization 
devotes 78\% of the integrated luminosity to the ``signal'' helicity 
configuration ($\rm{e}_L^-$, $\rm{e}_R^+$), 17\% to the ``background'' helicity 
combination ($\rm{e}_R^-$, $\rm{e}_L^+$) and 5\% equally shared amongst 
the polarization constraining like-sign helicity configurations 
of ($\rm{e}_L^-$, $\rm{e}_L^+$) and ($\rm{e}_R^-$, $\rm{e}_R^+$). The optimization was done assuming 90\% electron beam polarization 
and 60\% positron beam polarization.
The center-of-mass energies used in the 6-point scan 
are (160.6, 161.2, 161.4, 161.6, 162.2, 170.0) GeV with integrated luminosities in the ratios of 1:5:5:5:1:6 
respectively.
The current scan is optimized for measuring $M_W$.
There is room for further optimization and alternative strategies. 
Alternative scans better suited to measuring $\Gamma_W$ can also be envisaged.

The results from an ensemble of 1000 toy experiments is found to be an overall uncertainty on $M_W$ of 3.94~MeV. 
An example of one of these fits is shown in Table~\ref{tab:Minuit}.
In order to assess the effective contribution of the various systematic effects 
to the overall W mass error, the fits are then re-run for 6 different alternative fits 
where the parameters encompassing individual sources of systematic error that 
are normally fitted for, are fixed to their correct model values. 
The standard fit and the 6 variations lead to 7 estimates of the W mass uncertainty 
from which the intrinsic statistical error, the systematics associated with each 
source of uncertainty, and the total systematic uncertainty are estimated.
Note that since all the parameters are fitted for, the overall uncertainty including systematics is statistical in nature 
and can be improved further with increased integrated luminosity. The results of these 15-parameter fits 
are shown in Table~\ref{tab:systerrors}.

\begin{table}[!htb]
\begin{center}
\begin{tabular}{c|l|c|c}
Fit type &    Uncertainty source      &    $\Delta M_W$ [MeV]   &   $\Delta M_W$ (syst.) [MeV]  \\ \hline
fixbkg   &    Background              &   3.20          &     2.30    \\
fixpol   &    Polarization  &   3.73          &     1.27    \\
fixeff   &    Efficiency    &   3.86          &     1.18    \\
fixlum   &    Luminosity    &   3.76          &     0.78    \\
fixALRB  &    $A_{\rm{LR}}^B$       &  3.86          &     0.80    \\  \hline
fixall   &    Statistical   &   2.43           &             \\   
         & Systematic       &                  &       3.10      \\
standard & Total Error      &   3.94           &             \\   \hline
\end{tabular}
\caption{Mass errors for various fits for example 100~$\rm{fb}^{-1}$ 6-point scan with (90\%, 60\%) beam polarizations}
\label{tab:systerrors}
\end{center}
\end{table} 

We have also looked into 15-parameter fits with only two scan points such as illustrated in Table~\ref{tab:illustrate2}. 
The data at $\sqrt{s}=170$~GeV are quite useful for constraining the normalization parameters of the measurement 
and is retained. Such fits reach a smaller overall uncertainty. However they are more model-dependent and 
do little to demonstrate qualitatively the kinematic dependence of the cross-section at threshold. 
It would seem reasonable to make sure that the data collected would have enough degrees of freedom 
to test for example the cross-section dependence on $\Gamma_W$. 
The overall results of these scans are reported in Table~\ref{tab:Prospects}.

\begin{table}[!htb]
\begin{center}
\begin{tabular}{l|r|c|c|r|r|r|r}
    $\sqrt{s}$ (GeV) &   L ($\rm{fb}^{-1}$)   &  $f$  & $\lambda_{\rm{e}^{-}} \lambda_{\rm{e}^{+}} $   &   $N_{ll}$ & $N_{lh}$   & $N_{hh}$ & $N_
{RR}$ \\ \hline 
     161.4 &     86.957 &     0.7111  & $-+$   &   63443  &  262469  &  283058  &16927120 \\ 
     161.4 &     86.957 &     0.2000  & $+-$   &     463  &    1736  &    3740  & 3270457 \\ 
     161.4 &     86.957 &     0.0444  & $++$   &     219  &     922  &    1023  &  233371 \\ 
     161.4 &     86.957 &     0.0444  & $--$   &     997  &    4043  &    4463  &  299399 \\ \hline

     170.0 &     13.043 &     0.7111  & $-+$   &   29299  &  121140  &  123460  & 2542743 \\ 
     170.0 &     13.043 &     0.2000  & $+-$   &     126  &     567  &     900  &  490497 \\ 
     170.0 &     13.043 &     0.0444  & $++$   &      92  &     454  &     404  &   35300 \\ 
     170.0 &     13.043 &     0.0444  & $--$   &     445  &    1905  &    1927  &   44740 \\ \hline
\end{tabular}
\caption{Illustrative example of the numbers of events in each channel for a re-optimized 
100~$\rm{fb}^{-1}$ 2-point ILC scan with 4 helicity configurations.}
\label{tab:illustrate2}
\end{center}
\end{table}

\begin{table}[!htb]
\begin{center}
\begin{tabular}{c|c|c|c}
$|P(e^{-})|$ (\%) & $|P(e^{+})|$ (\%) & 100 fb$^{-1}$ & 500 fb$^{-1}$ \\ \hline
     80      &        30    &  6.02    &    2.88   \\
     90      &        30    &  5.24    &    2.60   \\
     80      &        60    &  4.05    &    2.21   \\
     90      &        60    &  3.77    &    2.12   \\ \hline
\end{tabular}
\caption{Total W mass uncertainty in MeV from polarized scan near threshold}
\label{tab:Prospects}
\end{center}
\end{table}

There are a number of issues that have not been treated in much depth including theoretical uncertainties, background composition, in 
particular four-fermion effects, and detailed modeling of event selection performance. 
However it is thought that the current treatment is appropriate for the current level of study. 

\newpage

\section*{Summary}
A threshold scan with polarized electron and positron beams can yield a precision measurement of $M_W$ at ILC. 
Errors at the few MeV level can be envisaged. With 100~fb$^{-1}$, and polarization values of (90\%, 60\%), 
the estimated uncertainty is 

\[ \Delta M_W (\rm{MeV}) = 2.4 \: \rm{(stat)} \oplus 3.1 \: (syst) \oplus 0.8 \: (\sqrt{s}) \oplus \rm{theory} \] 

The ILC design can and should evolve to make this feasible. Positron polarization is extremely helpful in 
controlling the polarization and background systematics. The highest polarization values would make 
such a measurement most impactful. Measurements with no positron polarization 
or even no electron polarization are obviously more challenging.

Eventual experimental precision approaching 2 MeV can be considered at ILC if one 
is able to dedicate 500 fb$^{-1}$ to such a measurement, and the physics perspective of the day demands it. 
Before embarking on such an extended run near threshold, one would certainly want to make sure that 
the center-of-mass energy systematic is indeed controlled at close to the envisaged level, 
that the theoretical uncertainties can be controlled adequately, and 
that such a program offers sufficient complementarity in the determination of the W mass to 
data already collected at higher center-of-mass energies synergistic with the main ILC program.


\Acknowledgments
I thank a number of people for their encouragement, comments or advice related to this measurement and surrounding issues 
including T. Barklow, P. Grannis, S. Heinemeyer, M. Hildreth, A. Kotwal, J. List, K. M\"{o}nig, G. Moortgat-Pick, M. Peskin, 
R. Tenchini, M. Thomson, E. Torrence, N. Walker.


\bibliographystyle{utphys}
\bibliography{eprint_lcws2015}

\end{document}